\documentclass[12pt]{iopart}
\usepackage[utf8]{inputenc}
\expandafter\let\csname equation*\endcsname\relax
\expandafter\let\csname endequation*\endcsname\relax
\usepackage{amsmath}
\usepackage{mathtools}
\usepackage{amsfonts}
\usepackage{feynmf}
\usepackage{dsfont}
\usepackage{bbm}
\usepackage{color}

\begin{document}

\title[Renormalizing Yukawa interactions in the Standard Model with matrices]{Renormalizing Yukawa interactions in the Standard Model with matrices and Noncommutative Geometry}

\author{Elliott Gesteau}

\address{Perimeter Institute for Theoretical Physics,\\Waterloo, Ontario N2L 2Y5, Canada}
\ead{egesteau@perimeterinstitute.ca}
\vspace{10pt}

\begin{abstract}
We show that gauge-independent terms in the one-loop and multi-loops $\beta$-functions of the Standard Model can be exactly computed from the Wetterich functional renormalization of a matrix model. Our framework is associated to the finite spectral triple underlying the computation of the Standard Model Lagrangian from the spectral action of Noncommutative Geometry. This matrix-Yukawa duality for the $\beta$-function is a first hint towards understanding the renormalization of the Noncommutative Standard Model conceptually, and provides a novel computational approach for multi-loop $\beta$-functions of particle physics models.
\end{abstract}

%\maketitle

\section{Introduction}
The Standard Model is thus far the most accurate model for particle physics, and manages to formulate electromagnetism and the weak and strong nuclear forces in one unified framework. Its successes are stunning, and it is fair to say that it is one of the cornerstones of theoretical physics in the twentieth century. However, its formulation heavily relies on the Lagrangian formalism of Quantum Field Theory, and a major challenge is to understand why its Lagrangian, which is quite lengthy and intricate, has such a form. One of the most successful attempts to tackle this problem is the Noncommutative Geometry approach, initiated by Connes and Chamseddine \cite{cc} in the late 90s. In this framework, the kinematics are encoded in an \textit{almost-commutative spectral triple}, while the dynamics are dictated by the Spectral Action principle, which, unlike a Lagrangian formulation, directly acts on the spectral triple geometry as a whole. The main object of the spectral triple is the Dirac operator, which is made of a continuous part and a finite part, the latter simply being a matrix.

\smallskip

Renormalizing the Standard Model is usually done through perturbative techniques, by the use of Feynman diagrams and counterterms. However, in addition to becoming very tricky at a high loop order, this technique heavily relies on the Lagrangian formulation, and therefore has no natural analogue in the noncommutative setting. Renormalization being a crucial feature of quantum field theories, it seems to be an important problem to understand what renormalization really means in the context of Noncommutative Geometry \cite{wvs1,wvs2}.

\smallskip

This paper is an attempt in such a direction. More precisely, we restrict our attention to the Yukawa sector of the Standard Model, whose noncommutative counterpart is encoded exclusively in the discrete part of the spectral triple. As a consequence, we may consider the Dirac operator as a matrix only. The right question to ask then is whether there exists a theory that one can build directly out of the finite Dirac operator, and which would yield the same renormalization equations as those of Yukawa theory. As the Dirac operator is a matrix, it is no surprise that the solution lies within the realm of matrix models. More specifically, we will construct an effective action involving a matrix variable, and renormalize it using the Wetterich equation \cite{frge,wetterich}. In such a context, renormalization is implemented directly within the path integral by the introduction of a regulator, thus making the effective action energy-dependent. The Wetterich equation then relates the energy dependence of the effective action to the regulator, and gives an ODE for the RG flow of the matrix model. 

\smallskip

In this paper, we prove that for an accurate choice of regulator, one can recover the one-loop $\beta$-function of some parameters in Yukawa theory without some charged Higgs terms, with the exact same coefficients, thanks to the Wetterich scheme applied to a matrix model. We also argue that our results extend to an arbitrary order in the loop expansion, with the exception of not encompassing contributions from the Higgs 4-vertex. Our model clears the meaning of the renormalization of a spectral action, and could have some applications for computational purposes. It seems to be a promising first step towards an alternative renormalization of the Standard Model, the ultimate goal being to find a (multi)-matrix model whose functional renormalization yields the same $\beta$-functions as the ones of perturbative Lagrangian theory, for all Standard Model parameters.

\smallskip

The paper is organized as follows: in Section \ref{sec:Yuk}, we describe Yukawa interactions and the Higgs mechanism in the usual Lagrangian Standard Model, and their renormalization. We then describe the noncommutative geometry of the Standard Model as well as the Spectral Action principle which encodes its dynamics, with a particular emphasis of the discrete part of the spectral triple, which corresponds to the Yukawa sector. In particular, we express the one-loop $\beta$-functions of the Yukawa parameters in terms of traces of even powers of the finite Dirac operator, in such a way that the equations are clearly reminiscent of matrix models. In Section \ref{sec:frg}, we introduce Wetterich's renormalization theory for matrix models. In Section \ref{sec:duality}, we use this framework to construct a matrix model whose effective action is inspired from the Spectral Action and exactly reproduces the previously stated one-loop $\beta$-function without its charged Higgs part, including coefficients. A nontrivial step in this construction is the choice of regulator, which has to reproduce the exact same energy-independent coefficients as in the Standard Model $\beta$-function in order to prove that the model provides insight into Yukawa theory, while still verifying physical asymptotic requirements. The existence of such a regulator which is consistent with our model is proven in the appendix of the paper. In Section \ref{sec:perspec}, we give some perspectives on our result. In particular, we give a precise argument why the method can be extended to an arbitrary loop order except for the contributions of Feynman diagrams including a Higgs 4-vertex. We also comment on the independence of loop order and truncation order in our context, and on how regulator universality could give more flexibility and computational power to our model. We also give a few ideas to construct a more realistic version of our renormalization duality: in close analogy to the spectral action principle, we explain how an even smearing function can create a model which incorporates the Higgs self-interaction as well as gauge couplings. More briefly, we speculate on how a multi-matrix model could allow to incorporate some charged Higgs terms.

\section{Yukawa interactions and the Noncommutative Standard Model}
\label{sec:Yuk}
\subsection{Yukawa interactions and the Higgs mechanism}

We consider here the minimal 
Standard Model extended with massive Dirac neutrinos. The Yukawa interactions couple the Higgs boson to the fermionic fields through a Lagrangian of the form 
\begin{equation}\mathcal{L}_\mathrm{Yuk}:=-Y_u^{\dagger ij}\bar{Q}_L^{i}\epsilon H^\ast u_R^{j} -Y_d^{\dagger ij}\bar{Q}_L^{i}H d_R^{j} -Y_\nu^{\dagger ij}\bar{L}_L^{i}\epsilon H^\ast\nu_R^{j} -Y_e^{\dagger ij}\bar{L}_L^{i} He_R^{j}+\mathrm{hc}.\end{equation}
Here $H$ is the Higgs boson field, the $Y$'s are the Yukawa matrices, $L$ and $Q$ are the lepton and quark doublets, $u$, $d$, $e$, $\nu$ the up, down, electron and neutrino singlets, and $\epsilon$ is an antisymmetric tensor acting on the left indices.\bigskip\\
In what follows, we shall be primarily interested in the renormalization of the Yukawa matrices. The $\beta$-functions for the running of those couplings at one loop, following the results of \cite{ackprw,beta}, take the form  
\begin{equation}\label{1}16\pi^2 \partial_t Y_\nu=Y_\nu\left(\frac{3}{2}Y_\nu^\dagger Y_\nu-\frac{3}{2}Y_e^\dagger Y_e+\mathrm{Tr}(Y_e^\dagger Y_e+Y_\nu^\dagger Y_\nu+3Y_d^\dagger Y_d+3Y_u^\dagger Y_u)+\mathrm{gauge\;terms}\right),\end{equation}\begin{equation}\label{2}16\pi^2 \partial_t Y_e=Y_e\left(\frac{3}{2}Y_e^\dagger Y_e-\frac{3}{2}Y_\nu^\dagger Y_\nu+\mathrm{Tr}(Y_e^\dagger Y_e+Y_\nu^\dagger Y_\nu+3Y_d^\dagger Y_d+3Y_u^\dagger Y_u)+\mathrm{gauge\;terms}\right),\end{equation}\begin{equation}\label{3}16\pi^2 \partial_t Y_d=Y_d\left(\frac{3}{2}Y_d^\dagger Y_d-\frac{3}{2}Y_u^\dagger Y_u+\mathrm{Tr}(Y_e^\dagger Y_e+Y_\nu^\dagger Y_\nu+3Y_d^\dagger Y_d+3Y_u^\dagger Y_u)+\mathrm{gauge\;terms}\right),\end{equation}\begin{equation}\label{4}16\pi^2 \partial_t Y_u=Y_u\left(\frac{3}{2}Y_u^\dagger Y_u-\frac{3}{2}Y_d^\dagger Y_d+\mathrm{Tr}(Y_e^\dagger Y_e+Y_\nu^\dagger Y_\nu+3Y_d^\dagger Y_d+3Y_u^\dagger Y_u)+\mathrm{gauge\;terms}\right),\end{equation}
in the variable $t=\log(\mu/\mu_0)$, with $\mu$ the energy scale.
At the electroweak phase transition, the theory undergoes spontaneous symmetry breaking and $H$ acquires a vacuum expectation value for its second component. The charged component is then cancelled by the gauge freedom, and there remains only a neutral, scalar Higgs field.

The $\beta$-function of the Standard Model at one, two, and three loops is discussed in \cite{ackprw}, \cite{beta2} and \cite{beta3}, with explicit computations, while to our knowledge no full explicit computation for higher loop order is available in the literature. One of the purposes of this paper is to show that the matrix model method presented here could provide a simpler path to the explicit computation of some higher loop contributions.

\subsection{The finite spectral triple of the Standard Model}

It is shown in \cite{ccm} (see also Chapter~1 of \cite{ncgqm}) that the full Lagrangian of the Standard Model extended with right-handed neutrinos and Majorana mass terms can be computed from the spectral action principle of \cite{cc}, to which we come back in the next subsection, applied to geometric fluctuations of an almost commutative space, which is the product of a four-dimensional smooth manifold and a discrete noncommutative space (a finite spectral triple). Over this finite geometry the Dirac operator $D$ is a finite-dimensional self-adjoint matrix. We recall here briefly the form of the finite spectral triple of \cite{ccm} and then we focus on the case without Majorana masses but with non-trivial Dirac masses and the resulting Yukawa terms.

\smallskip

A finite spectral triple is a datum $(A,H,D)$ of a finite-dimensional complex $C^*$-algebra $A$ acting
on a finite-dimensional Hilbert space $H$, together with a self-adjoint linear operator $D$ on $H$. A finite spectral triple  $(A, H, D)$ is even if there is a 
${\mathbb Z}/2{\mathbb Z}$-grading $\gamma$ on $H$ with
$\gamma^*=\gamma$, $\gamma^2=1$, $[\gamma,a]=0$ for all $a\in A$ and
$\gamma D + D \gamma =0$. 
Moreover, a finite spectral triple  $(A, H, D)$
has a real structure if there exists an anti-unitary operator $J: H \to H$
with the properties that $a^0 := J a^* J^{-1}$ defines a right action of $A$ on $H$
with $[ a, b^0]=0$, for all $a,b\in A$ and satisfying the ``order-one condition"
$[[D,a],b^0]=0$ for all $a,b\in A$. Moreover, the anti-unitary $J: H \to H$
should satisfy 
$$ J^2 = \varepsilon, \ \ \ \ JD = \varepsilon' DJ, \ \ \text{and} \ \
J\gamma = \varepsilon'' \gamma J $$
with $\epsilon, \epsilon',\epsilon'' \in \{\pm 1 \}$ where the third condition applies if the spectral triple is even.

\smallskip

The left-right symmetric spectral triple of \cite{ccm} has a real algebra $A_{LR,\mathbb{R}}=\mathbb{C}\oplus \mathbb{H}_L \oplus \mathbb{H}_R \oplus M_3(\mathbb{C})$, where $\mathbb{H}$ is the real algebra of quaternions, and Hilbert space $H$ is obtained by taking the sum $M$ of all irreducible odd bimodule for the left-right symmetric algebra, where the odd condition means that the involution $s=(1,-1,-1,1)$ in the algebra acts by $Ad(s)=-1$. (These are representations of the complex algebra $ B=\oplus^{4-times} M_2(\mathbb{C})\oplus M_6(\mathbb{C})$.) The Hilbert space is then given by a direct sum of three (the number of generations) copies of this bimodule $M$ of dimension $\dim_\mathbb{C} M = 32$. The bimodule $M$ decomposes into two parts (matter/antimatter sectors)
$M=E\oplus E^o$ interchanged by the real structure $J(\xi,\bar\eta)=(\eta,\bar\xi)$, while the grading is given by $\gamma =c - J c J$ with $c=(0,1,-1,0)\in A_{LR,{\mathbb R}}$, with $J^2=1$ and $J\gamma=-\gamma J$ (KO-dimension six).  
We refer the reader to \cite{ccm} and to Chapter~1 of \cite{ncgqm} for the details of this construction and the explicit identification of a basis of $H$ with the fermion fields of the Standard Model.
The main point that we need to recall here is the fact that the structure of the Dirac operator is determined by the self-adjointness and the order-one condition. In particular, it is shown in \cite{ccm} that if one requires that the Dirac operator $D$ intertwines the three copies of $E$ with the three copies of $E^o$ in the Hilbert space, then the the largest subalgebra on which the order-one condition can be satisfied is $A=\mathbb{C}\oplus \mathbb{H}\oplus M_3(\mathbb{C})$, with $\mathbb{C}$ embedded diagonally in the complex numbers and one of the two copies of the quaternions in $A_{LR,\mathbb{R}}$, breaking the left-right symmetry. The term in the Dirac operator intertwining $E$ and $E^o$ sectors corresponds to the Majorana mass terms, see \cite{ccm}. The general form of the Dirac operator is then shown to be (up to normalization) given by 
$$ D =\left(\begin{matrix}S &T^*\\
T &\bar S\end{matrix} \right) \ \  \text{ with } \ \ 
 S = S_\ell \, \oplus (S_q \otimes 1_3) $$ with $T$ the Majorana masses term, and with matrices 
 \begin{equation}\label{SellSq} S_\ell=\begin{bmatrix}0&0&Y_\nu^\dagger&0\\0&0&0&Y_e^\dagger\\Y_\nu&0&0&0\\0&Y_e&0&0\end{bmatrix} \ \ \text{ and } \ \  
 S_q=\begin{bmatrix}0&0&Y_u^\dagger&0\\0&0&0&Y_d^\dagger\\Y_u&0&0&0\\0&Y_d&0&0\end{bmatrix}, \end{equation}
 where $Y_\nu$, $Y_e$, $Y_u$, $Y_d$ are, respectively, the Yukawa terms for the neutrinos, charged leptons, u/c/t  and d/s/b quarks. These include the masses and the Cabibbo--Kobayashi--Maskawa and Pontecorvo--Maki--Nakagawa--Sakata mixing matrices, for the quarks and lepton sectors, respectively. In the following, we consider this same finite noncommutative geometry, in the absence of the intertwining term $T$.

\subsection{The Spectral Action}

In the Noncommutative Standard Model, the dynamics of the theory are encoded in the Spectral Action principle applied to an \textit{inner fluctuation} of the Dirac operator, which is an alternative to the Lagrangian formulation. An inner fluctuation of the Dirac operator $D$ is an operator:\begin{equation}D_A:=D+A+JAJ^{-1},\end{equation}where $A$ is Hermitian of the form:\begin{equation}A=\sum_ia_i[b_i,D].\end{equation}where the $a_i$ and $b_i$ are in $\mathcal{A}$. Then, the Spectral Action has two parts, a bosonic one and a fermionic one. The fermionic part involves subtle properties of the real structure, and we shall not need it for our purposes, therefore we refer the reader to \cite{ncgqm} for more details. The bosonic part has the following form: \begin{equation}S:=\mathrm{Tr}f\left(\frac{D_A}{\Lambda}\right),
\end{equation}where $f$ is an even smearing function and $\Lambda$ a cutoff. In \cite{ncgqm}, the authors prove that the Spectral Action principle allows to rederive the Lagrangian of the Standard model with massive neutrinos and Majorana mixing, coupled to gravity. An important point here will be that as we saw, the Yukawa couplings are encoded solely in the matrix $D$ which is the finite part of the Dirac operator, while gauge couplings at unification and the Higgs self-coupling are also encoded in the smearing function $f$, giving them a fundamentally different nature in the noncommutative context.

\smallskip

If $f$ is analytic, the spectral action can therefore be decomposed in even monomials in $D_A$. This property will be crucial for the construction of our theory space. There is a long story behind the theory of the Spectral Action (\cite{cc}, \cite{ccm}, \cite{ncgqm}, \cite{cosmo}), and although we will only merely touch upon its surface here, it is important to bear in mind for what follows that the space of spectral actions is a very natural setup to study gravity theories on noncommutative spaces (a particular case of which is the Noncommutative Standard Model.) 

\smallskip

In the case of a finite spectral triple having a matrix Dirac operator, the spectral action just turns into a matrix model action on the inner fluctuation matrix $A+JAJ^{-1}$. This is a first hint of the matrix model we will construct later.

\subsection{Renormalization of the finite geometry}

We now come back to the finite Dirac operator of the Standard Model, and first consider the explicit form of the traces of even powers of the Dirac operator of the finite spectral triple.
In the absence of Majorana masses for neutrinos, $D$ is constituted of blocks of $3\times 3$  matrices, which are the Yukawa couplings of the theory, in the form 
\begin{equation}D=\mathrm{diag}(S_\ell,S_q,S_q,S_q,\overline{S_\ell},\overline{S_q},\overline{S_q},\overline{S_q}).\end{equation}
with $S_\ell$ and $S_q$ as in \eqref{SellSq}.
The traces of powers of the Dirac operator are therefore of the form \begin{equation}\label{TrD2n}
\frac{1}{4}\mathrm{Tr}(D^{2n})=\mathrm{Tr}((Y_e^\dagger Y_e)^n)+\mathrm{Tr}((Y_\nu^\dagger Y_\nu)^n)+3\mathrm{Tr}((Y_u^\dagger Y_u)^n)+3\mathrm{Tr}((Y_d^\dagger Y_d)^n).\end{equation}

\smallskip

A natural question to ask is then how one can directly renormalize the Dirac operator. We consider the one-loop $\beta$-function of the Standard Model, as it is usually computed via perturbative quantum field theory, and we reinterpret it as an equation for the Dirac operator of the finite spectral triple.

\smallskip

We now derive the one-loop $\beta$-functions of the traces of even powers of the Dirac operator appearing in equation \eqref{TrD2n}. Note that by cyclicity of the trace,
\begin{align}16\pi^2\partial_t\mathrm{Tr}((Y_e^\dagger Y_e)^n)&=n16 \pi^2\mathrm{Tr}(\partial_t(Y_e^\dagger Y_e)(Y_e^\dagger Y_e)^{n-1})\\&=n16\pi^2\mathrm{Tr}(\partial_tY_e^\dagger Y_e(Y_e^\dagger Y_e)^{n-1})+n16\pi^2\mathrm{Tr}(Y_e^\dagger \partial_tY_e(Y_e^\dagger Y_e)^{n-1}).\end{align}
Now using equation \eqref{2} for the derivative of $Y_e$ alongside with cyclicity of the trace, we get 
\begin{multline}16\pi^2\partial_t\mathrm{Tr}((Y_e^\dagger Y_e)^n)=3n\mathrm{Tr}((Y_e^\dagger Y_e)^{n+1})-3n\mathrm{Tr}((Y_e^\dagger Y_e)^{n}Y_\nu^\dagger Y_\nu)+\frac{n}{2}\mathrm{Tr}(D^2)\mathrm{Tr}((Y_e^\dagger Y_e)^{n})\\+\mathrm{gauge\; terms}.\end{multline}
We can then sum over all four families of Yukawa couplings and conclude that at one loop, the $\beta$-functions of its successive powers can be expressed as 
\begin{equation}\label{reneqTr}
16\pi^2\partial_t\mathrm{Tr}(D^{2n})=3n\mathrm{Tr}(D^{2n+2})+\frac{n}{2}\mathrm{Tr}(D^2)\mathrm{Tr}(D^{2n})+\mathrm{charged\;Higgs\; term}+\mathrm{gauge\; terms}.\end{equation}

If the Yukawa couplings are chosen to be Hermitian, which is always possible up to the equivalence relation we will describe below, the charged Higgs term takes the simple form
\begin{equation}\mathrm{charged\;Higgs\; term}=-3n\mathrm{Tr}(D^{2n}\widetilde{D}^2),\end{equation}
where $\widetilde{D}$ is the twisted Dirac operator, which is obtained by switching respectively the up and down, and electron and neutrino couplings. 

\smallskip

The main idea here is that all gauge independent terms in the one loop $\beta$-functions of traces of even powers of the Dirac operator can be expressed in terms of traces of even powers of the Dirac operator, and possibly of $\widetilde{D}$. Note that information about these successive traces is enough to know the eigenvalues of the square of the Dirac operator, which are proportional to the fermion masses.

\smallskip

The three terms in \eqref{reneqTr} have different origins in the renormalization process. The term involving $\mathrm{Tr}(D^2)$ comes from the fermion loop correction to the Higgs propagator, and the term involving the twist comes from charged Higgs interactions.

\smallskip

In \cite{ccm}, \cite{ncgqm} it was shown that the possible choices of Dirac operators for the finite noncommutative geometry underlying the Standard Model with right-handed neutrinos and Majorana mass terms are parameterized by a moduli space given by a product ${\mathcal C}_3 \times {\mathcal C}_1$,
where ${\mathcal C}_3$ consists of pairs $(Y_d,Y_u)$ modulo the equivalence 
$Y'_d=W_1\,Y_d\,W_3^*$ and $Y'_u=W_2\,Y_u\,W_3^*$
implemented by unitary matrices $W_j$, so that
\begin{equation}\label{GL3quot}
 {\mathcal C}_3 = (U(3)\times U(3))\backslash
{\rm GL}_2({\mathbb C}) \times {\rm GL}_2({\mathbb C})/ U(3). 
\end{equation}
This part of the moduli space of finite Dirac operators accounts for all the Yukawa parameters of the quark sector. The lepton part ${\mathcal C}_1$ is a fibration over another copy of ${\mathcal C}_3$, which accounts for the Yukawa terms for the leptons including Dirac neutrino masses and a Pontecorvo--Maki--Nakagawa--Sakata mixing matrix, and with fiber the set of symmetric matrices $Y_R$ modulo the scaling equivalence 
$Y_R \mapsto \lambda^2 Y_R$, which accounts for the Majorana terms. If we consider only the extension of the minimal standard model by Dirac neutrino masses without the Majorana terms, then the moduli space of the finite Dirac operators consists of a product of two copies of the quotient of \eqref{GL3quot}. 

\smallskip

The flow equation for the Dirac operator obtained in \eqref{reneqTr} from the one loop $\beta$-function of the standard model with Dirac neutrino masses can be interpreted as resulting from a flow on the moduli space of Dirac operators. Indeed, the fact that the flow is expressed in terms of traces of powers of the Yukawa matrices in such a way ensures that it is completely determined by invariants of the equivalence relation of \eqref{GL3quot}.

\section{Nonperturbative renormalization and the Functional Renormalization Group}
\label{sec:frg}
Now that we have formulated the $\beta$-functions of Yukawa matrices in terms of the Dirac operator, we may move on to constructing a dual matrix model, which is the main purpose of this paper. As we will use functional renormalization for this paper, we now briefly review its setting \cite{frge,wetterich}, adapted to our context.
Since the results are about fixed size matrix models, we will present the theory in such a context, which has, to our knowledge, has never been done explicitly in the literature. For simplicity, we will assume that all matrices we use are real symmetric.

\smallskip

Let us consider a matrix field theory described by a generating functional 
\begin{equation}Z[J]=\int D[\tilde{A}]e^{-S[\tilde{A}]+\mathrm{Tr}(J\tilde{A})},
\end{equation}
where the integration variable $\tilde{A}$ is a finite size matrix of fixed dimensions and $J$ is a matrix of the same size. Note that this is the exact analogue of a quantum field theoretic path integral, for which Functional Renormalization is more frequently used, except here there is a finite number of degrees of freedom in the theory, because fields are matrices.

\smallskip

The idea is to introduce an additional energy scale $t$ dependent term in the exponential to account for the effects of renormalization and cancel the IR divergences of the model. This term has the form \begin{equation}\Delta S_t[\tilde{A}]=\frac{1}{2}\mathrm{Tr}(\tilde{A}R_t\tilde{A}),\end{equation}
where $R_t$, called the regulator, has the size of the tensor product of $A$ with itself. It depends in general on the entry $q$, and it accounts for the effects of renormalization. It is chosen freely except for the three following properties:
\begin{equation}\label{12}\underset{t \to 0}{\lim}R_t(q)=0.\end{equation}\begin{equation}\label{13}\underset{t \to \infty}{\lim}R_t(q)=\infty.\end{equation}\begin{equation}\label{14}\underset{\frac{q}{t} \to 0}{\lim}R_t(q)>0.\end{equation}
These properties have a precise meaning. Namely, \eqref{12} ensures that the flowing action approaches the usual effective action in the limit where $t$ goes to 0, while \eqref{13} ensures that the theory has a classical limit. Finally, \eqref{14} shows that $R_t(q)$ behaves as an IR regulator.

\smallskip

The energy dependent generating functional $Z_t[J]$ then becomes
\begin{equation}Z_t[J]=e^{-W_t[J]}=\int D[\tilde{A}]e^{-S[\tilde{A}]-\Delta S_t[\tilde{A}]+\mathrm{Tr}(J\tilde{A})},
\end{equation}
where
$W_t[J]$ is the energy scale dependent connected generating function. It then follows that
\begin{equation}\label{id2}\partial_t W_t=-\frac{1}{2}\mathrm{Tr}\left((\partial_t R_t) W_t^{(2)}\right)-\frac{1}{2}\mathrm{Tr}(A(\partial_t R_t) A).\end{equation}
The background field $A$ is defined  by
\begin{equation}A_{ij}=\frac{\partial W_t[J]}{\partial J_{ij}},
\end{equation} 
and the energy scale dependent effective action as a modified Legendre transform of $W_t[J]$ is given by
\begin{equation}\label{gamma}\Gamma_t[A]:=\mathrm{Tr}(JA)-W_t[J]-\Delta S_t[A].
\end{equation}
%The identity
%\begin{equation}\delta_{ij}=\sum_k\left(\frac{\partial{A}}{\partial{J}}\right)_{ik}\left(\frac{\partial{J}}{\partial{A}}\right)_{kj}
%\end{equation} implies that
An application of the chain rule then yields that
\begin{equation}\label{id1}W_t^{(2)}=(\Gamma_t^{(2)}+R_t)^{-1}.\end{equation}
Deriving \eqref{gamma} and using identities \eqref{id1} and \eqref{id2} %\Note{[Please always label the equations are refer to labels!!]}
yields the Wetterich equation, which describes how the effective action of the theory flows with the energy scale. We have
\begin{equation}\partial_t\Gamma_t=\frac{1}{2}\mathrm{Tr}\left(\frac{\partial_t R_t}{R_t+\Gamma_t^{(2)}}\right).
\end{equation}
The Wetterich equation is a powerful nonperturbative tool to renormalize a quantum field theory. In the next section we will see how to renormalize the Yukawa parameters by applying the Wetterich equation to a matrix model.

\section{A matrix-Yukawa duality for $\beta$-functions}
\label{sec:duality}
In this section, we derive the main result of the paper, which proves that the one-loop Yukawa $\beta$-functions can be derived from the Wetterich renormalization of a matrix model, with the exact same coefficients.

\subsection{Objectives and overview of the method}

\eqref{reneqTr} seems very close to some equations obtained in the context of matrix models. This is a first hint that there should exist a matrix model whose Wetterich renormalization yields the same equation. The objective of the upcoming construction is to realize this claim. In other words, we want to construct an effective action for a matrix variable in a clever way, such that it will generate \eqref{reneqTr} when renormalized. Here, we outline the main steps of our construction.

\smallskip

The first step is to construct the theory space, which will describe the general form of the effective action. We will propose such a construction, and justify it by some remarks on Noncommutative Geometry and the Spectral Action. Then, we will simplify the Wetterich equation in the case of a weak contribution of interaction terms in our model. We will then give a particular choice of theory (which in particular allows only Yukawa couplings to run), and argue that in such a context, the existence of an effective action which reproduces the one-loop $\beta$-function of Yukawa interactions (without the charged Higgs and gauge parts) can be reduced to a system of ODE's on the regulator. Under the assumption of the existence a physically meaningful solution of the system, which is proven in the appendix, we derive our final result.

\subsection{The theory space}

We construct our theory space in close analogy with the Spectral Action principle of Noncommutative Geometry. In Noncommutative Geometry, as briefly stated in subsection 2.3, the dynamics of the theory are encoded by applying the Spectral Action principle to a perturbed Dirac operator $D_A$, which depends on the perturbation matrix $A$. The general form of the Spectral Action being \begin{equation}S:=\mathrm{Tr}f\left(\frac{D+A+JAJ^{-1}}{\Lambda}\right),
\end{equation}where $f$ is an even smearing function, the expansion of a spectral action of a perturbed Dirac operator will contain terms of the form $\alpha\mathrm{Tr}(M(D,A+JAJ^{-1}))$ where $M$ is a unit coefficient monomial with even degrees both in $D$ and $A+JAJ^{-1}$, and $\alpha$ is a coefficient which depends on $f$. It is therefore natural to consider a theory space made of effective actions of the form \begin{equation}\Gamma[A]=\sum_{n,k\in\mathbb{N}}a_{n,k}M_{2k,2n}(D,A),\end{equation}where we have substituted $A+JAJ^{-1}$ by $A$ and $M_{2k,2n}(D,A)$ is the trace of a polynomial made of monomials of partial degrees $2k$ and $2n$ respectively in $D$ and $A$. The couplings that can vary a priori are the $a_{n,k}$ and the matrix $D$. However, in our case, Yukawa interactions are encoded solely within the matrix $D$. On the contrary, the $a_{n,k}$ would influence gauge couplings and the Higgs self-coupling, as they would correspond to the even smearing function $f$ of the Spectral Action. Therefore, in what follows, we will be interested in constructing the full renormalization flow in theory space in subspaces where almost all $a_{n,k}$ are constant, so that our renormalization equations will only involve the Dirac operator, as it is the case in \eqref{reneqTr}. The cost will be a technical construction of the regulator. We will return to the very interesting question of how varying the $a_{n,k}$ can enrich our model in the last section.

\subsection{Simplifying the Wetterich equation}

Now, we take a particular choice of effective action within theory space for which non quadratic terms in $A$ are small. In such a context, we show that it is possible to simplify the Wetterich equation in the case of a diagonal regulator. Consider a theory of with an effective action  
\begin{equation}\Gamma[A]:=Y+\frac{Z}{2}\mathrm{Tr}\left(A^2\right)+\eta \sum\limits_{n=2}^\infty M_{2n}(D,A),
\end{equation} 
where $D$ is the Dirac operator of the previously discussed discrete geometry, $A$ is a matrix of the same size which we consider as our field, and $M_{2n}$ is the trace of an even polynomial in $D$ and $A$, made of monomials of partial degree $2n$ in $A$. We will consider all matrices of the model to be real symmetric (which amounts to neglecting CP violation in the Dirac operator) in order to simplify the calculation, which can then be extended to a more general complex case following the rules of \cite{Benedetti}.

\smallskip

The Wetterich equation applied to that effective action gives 
\begin{equation}\partial_t\left(Y+\frac{Z}{2}\mathrm{Tr}(A^2)+\eta\sum\limits_{n=2}^\infty M_{2n}(D,A)\right)=\frac{1}{2}\mathrm{Tr}\left(\frac{\partial_t R_t}{(R_t+Z)\mathds{1}+\eta\sum\limits_{n=2}^\infty M_{2n}(D,A)^{(2)}}\right).\end{equation}

We now choose our regulator $R_t\mathds{1}$ (where $\mathds{1}$ is the superidentity in the space of matrices $A$) in such a way that we have
\begin{equation}\label{regeq}
R_t+Z=C.
\end{equation} 
where $C$ is a scalar function which will be specified later. A perturbative expansion at first order in $\eta$ can then be performed, giving 
\begin{equation}\partial_t\left(Y+\frac{Z}{2}\mathrm{Tr}(A^2)+\eta\sum\limits_{n=2}^\infty M_{2n}(D,A)\right)=\frac{1}{2C}(\partial_t R_t)\mathrm{Tr}\left(\mathds{1}-\frac{\eta}{C}\sum\limits_{n=2}^\infty M_{2n}(D,A)^{(2)}\right).
\end{equation}
In what follows we will use the following identification ansatz: first set $A$ to $\epsilon\mathds{1}$, and then identify the terms with the same power dependence in $\epsilon$.

\subsection{Recovering the Yukawa $\beta$-function}

We now apply this method to a specific theory. We define the effective action as \begin{equation}\label{theory}\Gamma[A]:=Y+\frac{Z}{2}\mathrm{Tr}(A^2)+\eta\left(\sum\limits_{n=2}^\infty a_{2n}\mathrm{Tr}(D^{2n-2}A^{2n})+b_{2n}\mathrm{Tr}(D^2A^2D^{2n-4}A^{2n-2})\right).\end{equation}
The reason for this choice is that if we want to have a chance to reproduce the Yukawa $\beta$-functions, we want Wetterich theory to yield equations involving terms of the forms $\mathrm{Tr}(D^{2n})$ and $\mathrm{Tr}(D^{2})\mathrm{Tr}(D^{2n})$, according to \eqref{reneqTr}. Moreover, in order to have equations which just concern the Dirac operator (ie just Yukawa couplings and not gauge or Higgs couplings), we want the coefficients $a_{2n}$ and $b_{2n}$ to be constant, which is a priori not ensured by the setup. However, we shall see that a careful choice of regulator should ensure this property. We return to the question of varying $a_{2n}$ and $b_{2n}$ at the end of the paper.

\smallskip

The simplified Wetterich equation applied to $A=\epsilon Id$ then reads
\begin{multline}\partial_t\left(Y+\frac{Z}{2}\epsilon^2\mathrm{Tr}(Id)+\eta\left(\sum\limits_{n=2}^\infty \epsilon^{2n}(a_{2n}+b_{2n})\mathrm{Tr}(D^{2n-2})\right)\right)=\\\frac{1}{2C}(\partial_tR_t)(\mathrm{Tr}(\mathds{1})-\frac{\eta}{C}\sum\limits_{n=2}^\infty\epsilon^{2n-2}( n(2n-1)(1+\mathrm{Tr}(Id))a_{2n}\mathrm{Tr}(D^{2n-2})+\\4(n-1)b_{2n}\mathrm{Tr}(D^2)\mathrm{Tr}(D^{2n-4})+\\((n(2n-1)-4(n-1))(1+\mathrm{Tr}(Id))+4(n-1))b_{2n}\mathrm{Tr}(D^{2n-2})).\end{multline}
We then use our ansatz to project the equation on each power of $\epsilon$.\footnote{We leave a more detailed study of this projection rule in the framework of functional renormalization to future work.} 

\smallskip

We obtain the set of equations 
\begin{equation}\label{dtY}
\partial_t Y=\frac{\mathrm{Tr}(\mathds{1})}{2C}\partial_tR_t.
\end{equation}
\begin{equation}\label{dtZ}
\mathrm{Tr}(Id)\partial_t Z=-\frac{6\eta\partial_t R_t}{C^2}(1+\mathrm{Tr}(Id))(a_{4}+b_4)
\mathrm{Tr}(D^2).
\end{equation}
and for $n\geq2$ the equations
\begin{multline}\label{dtTr}
(a_{2n}+b_{2n})\partial_t(\mathrm{Tr}(D^{2n-2}))=\\-\frac{1}{2C^2}(\partial_t{R_t})(((n+1)(2n+1)(1+\mathrm{Tr}(Id))a_{2n+2}+(((n+1)(2n+1)-4n)\\\cdot(1+\mathrm{Tr}(Id))+4n)b_{2n+2})\mathrm{Tr}(D^{2n})+4nb_{2n+2}\mathrm{Tr}(D^2)\mathrm{Tr}(D^{2n-2})).\end{multline}

\smallskip

One now easily notices that \eqref{dtTr} looks extremely close to \eqref{reneqTr}. One is therefore tempted to define an induction relation between the $a_{2n}$ and $b_{2n}$ such that the same coefficients are respected. However, in order to derive this simple equation, we have supposed that they are independent of the energy scale, and in order for this to be true and at the same time recover \eqref{reneqTr}, whose coefficients are also independent of the energy scale, we need to ensure that
\begin{equation}\label{TrRCK}
\frac{\partial_tR_t}{C^2}=K,
\end{equation} 
where $K$ is independent of the energy scale. Now, it is far from obvious that satisfying \eqref{regeq}, \eqref{dtZ} and \eqref{TrRCK}
simultaneously, while still verifying the asymptotic conditions \eqref{12}, \eqref{13} and \eqref{14} is possible. This is important for our purpose, and it is fortunate that it indeed is. The proof of such a result is quite technical and explained in the appendix of this paper. For now, let us assume \eqref{TrRCK} and return to our initial problem. If we define the $a_{2n}$ and $b_{2n}$ recursively by 
\footnotesize
\begin{multline}
    -\frac{K}{2}\frac{(n+1)(2n+1)(1+\mathrm{Tr}(Id))a_{2n+2}+(((n+1)(2n+1)-4n)\cdot(1+\mathrm{Tr}(Id))+4n)b_{2n+2}}{a_{2n}+b_{2n}}\\=\frac{3(n-1)}{16\pi^2}, \end{multline}\normalsize \begin{equation}-\frac{K}{2}\frac{4nb_{2n+2}}{a_{2n}+b_{2n}}=\frac{n-1}{32\pi^2},\end{equation} we have for $n\geq1$ the expression
\begin{equation}\label{result}16\pi^2\partial_t\mathrm{Tr}(D^{2n})=3n\mathrm{Tr}(D^{2n+2})+\frac{n}{2}\mathrm{Tr}(D^2)\mathrm{Tr}(D^{2n}),\end{equation} which recovers the expression of the Yukawa $\beta$-function without charged Higgs and gauge terms, with the exact same coefficients.

\smallskip

As a quick summary, using the spectral action principle as a guideline, we constructed a matrix model which is completely independent from any particle physics consideration, and is dual to one-loop Yukawa theory, in the sense that it reproduces \eqref{reneqTr} (without its charged Higgs and gauge part) with the exact same coefficients. Let us stress the physical significance of \eqref{reneqTr} once more: as the even powers of the Dirac operator have eigenvalues proportional to even powers of the Yukawa masses, knowing their traces amounts to knowing the sums of all powers of these eigenvalues, and hence to reconstruct them. We will now show that this method can almost as successfully be generalized to account for an arbitrary loop order in the Feynman diagram approach.

\section{Perspectives}
\label{sec:perspec}
In this final section, we give some perspectives on our result and suggest some future research directions.

\subsection{Adding more loops}

The previous matrix model is dual to one-loop Yukawa theory. A natural question to ask is whether a similar construction can be made for an arbitrary loop order, which is the problem we now turn to.

\smallskip

In order to find, with any given coefficients, a $\beta$-function for the Dirac operator that only involves traces of its even powers, in analogy with the two first terms of \eqref{reneqTr} it is quite straightforward to see that it is sufficient to have equations for the Yukawa matrices $Y_i$ that are of the form \begin{equation}\label{standard}\partial_tY_i=Y_if\left(Y_i^\dagger Y_i,\{\mathrm{Tr}(D^{2n})\}_{n\in\mathbb{N}}\right),\end{equation}which is a generalization of the structure of equations \eqref{1}, \eqref{2}, \eqref{3} and \eqref{4}.
In \cite{beta2} and \cite{beta3}, the equations of the two and three-loop $\beta$-functions are of this form, up to the presence of the Higgs self-coupling and some charged Higgs contributions. In this section, we give some arguments to show that this is in fact true for all loop orders: a dual matrix model can be constructed for the renormalization of Yukawa interactions at an arbitrary number of loops, up to diagrams which involve a four Higgs vertex.

\smallskip

The counterterms to the Yukawa matrices come from the renormalization of the fermion-antifermion-Higgs vertex, which can be computed at a fixed loop order once the fermionic and bosonic two-point functions are already renormalized. A diagram will yield a counterterm of the form of the right hand side of \eqref{standard} if it contains an odd number of vertices involving $Y_i$, where $i$ is the type of the incoming fermion, and an even number of vertices involving other fermions. 

\smallskip

Apart from some charged Higgs contributions, the vertices involving other types of fermions only appear in fermion loops, and exactly yield traces of powers of the Dirac operator, as Yukawa matrices are traced over and summed with the right color factors. Therefore, the only diagrams which could pose a problem to generalize our matrix duality are diagrams containing fermion loops with an odd number of propagators. However, such diagrams cannot exist because of the chiral nature of Yukawa interactions in the Standard Model: a Yukawa vertex always couples a right-handed fermion to a left-handed fermion. As a result, all counterterms not involving a Higgs 4-vertex will have the right form, and our model can be extended to an arbitrary loop level.

\smallskip

\begin{fmffile}{graph}
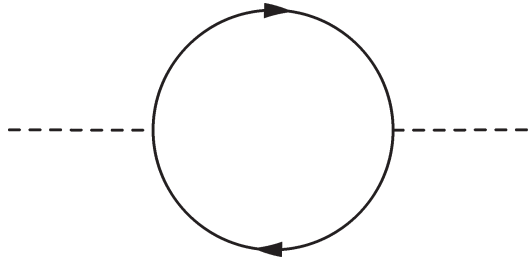
\begin{figure}
    \centering
    \begin{fmfgraph}(200,100)
    \fmfleft{i1}
    \fmfright{o1}
    \fmf{dashes}{i1,v1}
    \fmf{fermion,left,tension=.3}{v1,v3,v1}
    \fmf{dashes}{v3,o1}
    \end{fmfgraph}
    \caption{Fermion loops allow for traces of powers of the Dirac operator to appear.}
    \label{fig:my_label}
\end{figure}
\begin{figure}
    \centering
\begin{fmfgraph}(200,100)
\fmfleft{i1}
\fmfright{o1,o2}
\fmf{dashes,tension=.5}{i1,v1}
\fmf{fermion,tension=.1}{v1,v2,v3,v1}
\fmf{dashes,tension=.3}{v2,v4}
\fmf{dashes,tension=.3}{v3,v5}
\fmf{fermion}{o1,v4}
\fmf{fermion,tension=.2}{v4,v5}
\fmf{fermion}{v5,o2}
\end{fmfgraph}
    \caption{Odd fermion loops are forbidden by the chiral nature of the theory.}
    \label{fig:my_label}
\end{figure}
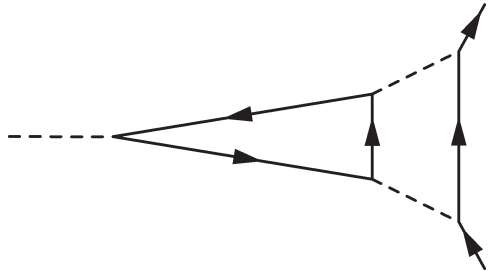

\smallskip

However, notice that this scheme does not encompass the renormalization of the Higgs four-point function, for which a quartic Higgs self-coupling needs to be introduced. At one loop, the Yukawa matrices corrections do not depend on it, but that result is no longer true for more loops. Therefore, a renormalization scheme for the Yukawa matrices which also encompasses the Higgs four-point function should have a more complicated multiloop structure. We will return to this problem in the last subsection.

\begin{figure}
    \centering
\begin{fmfgraph}(200,100)
\fmfleft{i1,i2}
\fmfright{o1,o2}
\fmf{dashes}{i1,v1}
\fmf{dashes}{i2,v2}
\fmf{fermion}{v1,v2,v4,v3,v1}
\fmf{dashes}{v3,o1}
\fmf{dashes}{v4,o2}
\end{fmfgraph}
    \caption{This divergence in the Higgs four-point function needs a quartic self-coupling to be cancelled.}
    \label{fig:my_label}
\end{figure}
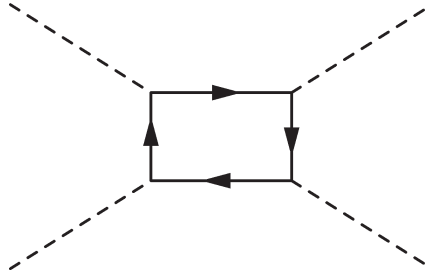
\begin{figure}
    \centering
    \begin{fmfgraph}(200,100)
    \fmfleft{i1}
    \fmfright{o1}
    \fmf{dashes}{i1,v1}
    \fmf{dashes,left,tension=.3}{v1,v3,v1}
    \fmf{dashes}{v1,v3}
    \fmf{dashes}{v3,o1}
    \end{fmfgraph}
    \caption{At two loops and more, the Higgs 4-vertex makes diagrams which are not taken into account appear in the matrix model.}
    \label{fig:my_label}
\end{figure}
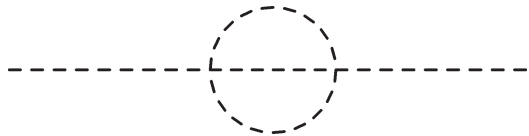
\end{fmffile}

\subsection{Loop orders vs truncations}

If we consider such models for higher order loops, their effective actions will, just like \eqref{theory}, be described by an infinite sum of traces of terms of the form $M_{n,k}(D,A)$, which are monomials of degree $2n$ and $2k$ in $D$ and $A$. The usual way that RG flows are actually calculated is through a truncation procedure, which approximates the effective action by truncating it at a finite order in $A$. Here, the interesting thing is that the truncation order is completely independent from the loop order: at every loop, $k$ will go from $0$ to $\infty$. Therefore, once the matrix model is constructed, the efficiency of the method is completely blind to the loop order, and a truncation at a given order should work equally well independently of the number of loops. This could be interesting for computational purposes, which we leave for future research.

\subsection{Physical interpretation and regulator universality}

Another interesting point is that as we previously discussed, the effective action $\Gamma[A]$ of our matrix model has the form of the spectral action of a matrix variable, whose degrees of freedom live in the finite noncommutative dimensions of spacetime (i.e. Higgs bosonic degrees of freedom if it is an inner fluctuation). This space being finite, there is no divergent amplitude to renormalize just inside it. However, recall that the full algebra describing the Standard Model is of the form $$C^\infty(M)\otimes \mathcal{A}_F,$$ where $\mathcal{A}_F$ is the finite algebra of matrices that describes our matrix model. The energy scale $t$ in the matrix model, which has not been specified so far, then has a natural interpretation in terms of a cutoff scale corresponding to the regularization of the smooth spacetime part of the geometry, to which our matrix model is blind. One can then interpret our matrix model as the restriction of a larger spectral action model which describes more of the Standard Model and also encodes the smooth spacetime degrees of freedom, and the renormalization scale as coming from a regularization the smooth spacetime part of this larger model. In particular, it would be very nice to describe this larger model, and maybe eventually even the full noncommutative standard model of \cite{ncgqm} as the large $N$ limit of matrix models that contain our Yukawa model as a submodel, and to interpret $N$ as the renormalization scale.

The above argument tells us that the physics of our matrix model is very closely related to the original standard model. In particular, in an extension of the matrix model containing a spacetime renormalization scale, we expect that \eqref{reneqTr} can still be recovered through functional renormalization. Now, note that we used a quite complicated method in the appendix to be able to recover \eqref{reneqTr} with the exact same coefficients. One of the strengths of functional renormalization is regulator universality: the physics of the matrix model will likely not depend on the choice of regulator. Hence, such matrix model constructions would leave a lot of freedom in the choice of regulator, and much simpler choices than the ones of the appendix would then also be suitable choices to describe the renormalization of the Standard Model. In particular, a simple enough choice could lead to a computationally more efficient way of renormalizing the Standard Model.

\subsection{Link to quantum gravity models}

It is worth noting that the spectral action contains all geometric information needed to formulate electromagnetism, weak and strong interactions, and gravity altogether in the framework of Noncommutative Geometry. Consequently, a more conceptual understanding of its renormalization could lead to interesting new perspectives for several physically relevant geometric models, especially matrix models for quantum gravity. Some features of our matrix model are shared with the explored in \cite{azarfar,barrett,barrett2,alicia,matrixmod}, and it would be interesting to know whether those similarities can be made more precise.

\subsection{Enriching the model: towards gauge and Higgs couplings and charged Higgs contributions}

This model is only a first step in the functional renormalization of the Noncommutative Standard Model. Ultimately, we want it to incorporate the Higgs self-coupling, and gauge couplings. In order to do this, we shall once again use the Spectral Action principle as a guideline. In the Noncommutative Standard Model, the Higgs self-coupling comes from a term of the form\begin{equation}\lambda=\frac{f_0}{2\pi^2}\mathrm{Tr}(D^4).\end{equation} This term contains a factor $f_0$ which comes from the test function $f$ of the spectral action, and also appears in the gauge couplings at unification \cite{ncgqm}. The presence of this extra factor seems inevitable: without it, in the expansion of the spectral action \cite{ncgqm}, the Higgs self-coupling would only depend on the trace of an even power of the Dirac operator, whose renormalization equation is already determined by equation \eqref{result}. Therefore there would not be enough freedom in the model to be able to renormalize the Higgs self-coupling to cancel the divergences of the new vertex. In other words, to go further than just Yukawa matrices, it seems inevitable to renormalize the smearing function $f$ of the spectral action in addition to the finite part of the Dirac operator. In the context of our matrix model, this will amount to allowing the coefficients $a_{2n}$ and $b_{2n}$ to vary. This will of course complicate the discussion a lot, and add a lot more freedom to the framework. In the end, though, we expect this to be the right way to simultaneously renormalize more couplings than just the Yukawa matrices, and we leave to future research the question of determining how many more $\beta$-functions of Higgs and gauge couplings can be renormalized at the same time as the Yukawa ones in such a fashion.

\smallskip

A last problem we have not touched upon is how to incorporate the charged Higgs term of \eqref{reneqTr}, involving the Dirac operator $\tilde{D}$ with switched lepton and quark matrices, into the model. There seems to be no natural way to do so naturally within the framework. The easiest thing one could think of is to treat $\tilde{D}$ as an independent matrix, and to write an analogous multi-matrix model involving $D$ and $\tilde{D}$. Once again, we shall leave this problem to future research.

\ack

It is a pleasure to thank Matilde Marcolli for all the inspiring ideas she gave me, as well as for her help and encouragement on this paper. I also thank Vincent Chen, Diego Garcia and especially Alicia Castro for helpful discussions. Research at Perimeter Institute is supported by the Government of Canada through Industry
Canada and by the Province of Ontario through the
Ministry of Research and Innovation.

\appendix
\section{Construction of the regulator}

One can reasonably wonder whether satisfying \eqref{regeq}, \eqref{dtZ} and \eqref{TrRCK}
simultaneously is possible. By substituting \eqref{dtZ} and \eqref{TrRCK} into the logarithmic derivative of \eqref{regeq}, we get an ODE for $C$ of the form
\begin{equation}\label{dtC}
\mathrm{Tr}(Id)\partial_t C=-\frac{6\eta\partial_t R_t}{C^2}(1+\mathrm{Tr}(Id))(a_{4}+b_4)
\mathrm{Tr}(D^2)+\mathrm{Tr}(Id)KC^2.
\end{equation}
This is a particularly simple case of the standard Riccati equation 
\begin{equation}\label{Riccati}
 C^\prime = \alpha_0(t) + \alpha_1(t) C + \alpha_2(t) C^2, 
\end{equation}
where we have 
\begin{equation}\label{alphai}
\alpha_0=-\frac{6\eta K}{\mathrm{Tr}(Id)}(1+\mathrm{Tr}(Id))(a_{4}+b_4)
\mathrm{Tr}(D^2), \ \ \  \alpha_1 = 0, \ \ \  \alpha_2 = K ,
\end{equation}
where in our case $\alpha_2$ is independent of $t$.
We use the standard Riccati substitution 
$$ V(t) = C(t) K \ \ \ \text{ and } \ \ \   V =- \frac{U'}{U}, $$
which gives the equation
\begin{equation}\label{implicit}U^{\prime \prime} +K\alpha_0(t) U =0 .\end{equation}

Note that the Cauchy-Lipschitz theorem guarantees that such a $U(t)$ is globally defined, as linear functions are globally Lipschitz.
The solution of the equation is then given, as long as $U$ does not vanish, by \begin{equation}C(t)=-\frac{U^\prime(t)}{\alpha_2U(t)}.\end{equation}
Then, we choose $a_4$ and $b_4$ such that $\alpha_0(t)$ is negative, and $K$ such that $\alpha_2$ is positive. A few remarks: first, since \begin{equation}R_t=\int KC^2,\end{equation} we have proved that an $R_t$ that verifies the system is globally defined for a given choice of initial condition. Another important point is that no explicit formula can be given for the expression of $R_t$, as \eqref{implicit} cannot be solved explicitly. Therefore we will need some general arguments to check the asymptotic requirements, given a set of initial conditions.

\smallskip

We now impose our initial conditions:
\begin{equation}U^\prime(0)<0
\end{equation} and \begin{equation}U(0)<0.\end{equation} These two conditions imply that $U$ decreases (therefore does not vanish) and goes to $-\infty$. We will also work under the additional assumption that 
$\frac{t}{-\alpha_0(t)}$ is bounded at infinity, which seems reasonable as our Yukawa couplings are expected to blow up, or even have a Landau pole.

\smallskip

We now want the regulator to match the asymptotic requirements. Checking that it is the case reduces to showing that $C^2=\left(\frac{U^\prime}{U}\right)^2$ does not have a convergent integral. To show this we use an energy estimate. For $t_0>0$, the linear ODE on $U(t)$ gives 
\begin{equation}\int_{t_0}^tU^\prime(t)U^{\prime\prime}(t)\mathrm{d}t+\alpha_2\int_{t_0}^t\alpha_0(t)U(t)U^\prime(t)\mathrm{d}t=0.\end{equation}
Then we obtain
$$\frac{U^\prime(t)^2}{2}-\frac{U^\prime(t_0)^2}{2}=\alpha_2\int_{t_0}^t(-\alpha_0(s))U(s)U'(s)\mathrm{d}s.$$Our hypothesis on the asymptotic behavior of $-\alpha_0(t)$ then allows to write, for some $\beta>0$, 
$$\frac{U^\prime(t)^2}{2}-\frac{U^\prime(t_0)^2}{2}\geq\beta\int_{t_0}^tsU(s)U^\prime(s)\mathrm{d}s=\beta t\frac{U(t)^2}{2}-\beta t_0\frac{U(t_0)^2}{2}-\beta\int_{t_0}^t\frac{U(s)^2}{2}\mathrm{d}s.$$
Dividing everything by $U(t)^2$ gives $$\frac{U^\prime(t)^2}{2U(t)^2}\geq\frac{U^\prime(t_0)^2}{2U(t)^2}-\beta t_0\frac{U(t_0)^2}{2U(t)^2}+\beta \frac{t}{2}-\frac{\beta}{U(t)^2}\int_{t_0}^t\frac{U(s)^2}{2}\mathrm{d}s.$$
The two first terms go to $0$ at infinity, therefore for $t$ large enough they are greater than $-\frac{\beta t_0}{4}$. As $U^2$ is increasing, the last term is greater than $-\beta\frac{t-t_0}{2}$. We therefore finally get \begin{equation}\left(\frac{U'(t)}{U(t)}\right)^2\geq\frac{\beta t_0}{2},\end{equation} 
for large $t$, which shows that $C^2$ does not have a convergent integral, and therefore that $R_t=\int KC^2$ blows up at infinity. Choosing the integration constant wisely while integrating $KC(t)^2$ therefore allows one to set the limit at $0$ of the regulator to be $0$, and the third condition being obviously satisfied (since the regulator has no momentum dependence). We have thus proved that our choice of $R_t$ is suitable.


\end{document}